# A Pairwise T-Way Test Suite Generation Strategy Using Gravitational Search Algorithm

Khin Maung Htay
*Faculty of Electronic Engineering Technology*
*Universiti Malaysia Perlis (UniMAP)*
02600 Arau, Perlis, Malaysia
khin@studentmail.unimap.edu.my

Rozmie Razif Othman
*Faculty of Electronic Engineering Technology*
*Universiti Malaysia Perlis (UniMAP)*
02600 Arau, Perlis, Malaysia
rozmie@unimap.edu.my

Amiza Amir
*Intelligent Computing Division,*
*Advanced Communication Engineering (ACE) Center of Excellence*
*Universiti Malaysia Perlis (UniMAP)*
01000 Kangar, Perlis, Malaysia
amizaamir@unimap.edu.my

Hasneeza Liza Zakaria
*Intelligent Computing Division,*
*Advanced Communication Engineering (ACE) Center of Excellence*
*Universiti Malaysia Perlis (UniMAP)*
01000 Kangar, Perlis, Malaysia
hasneeza@unimap.edu.my

Nuraminah Ramli
*Faculty of Electronic Engineering Technology*
*Universiti Malaysia Perlis (UniMAP)*
02600 Arau, Perlis, Malaysia
nuraminah@unimap.edu.my

*Abstract—* Software faults are commonly occurred due to interactions between one or more input parameters in complex software systems. Software test design techniques can be implemented to ensure the quality of the developed software. Exhaustive testing tests all possible test configurations; however, it is infeasible considering time and resource constraints. Pairwise t-way testing is a sampling strategy that focuses on testing every pair of parameter combination, effectively reducing the generated test size as opposed to testing exhaustively. In this paper, we propose a new pairwise t-way strategy called Pairwise Gravitational Search Algorithm Strategy (PGSAS). PGSAS utilizes Gravitational Search Algorithm (GSA) for generating optimal pairwise test suites. The performance of PGSAS is benchmarked against existing t-way strategies in terms of test suite size. Preliminary results showcase that PGSAS provides competitive results in most configurations and outshines other strategies in some cases.

I. INTRODUCTION

Software testing plays a crucial role in the Software Development Lifecycle (SDLC). As the released software typically consists of many inputs with various configurations, software test design techniques are needed to produce quality test cases that can identify potential software faults. Interaction failures, which occur when the System Under Test (SUT) input parameters interact, are a common cause of software failures. While exhaustive testing is designed to assess all possible thorough list of test cases, it is, in fact, impractical due to the limitation of testing time and resources [1]. To tackle the issue of exhaustive testing, t-way testing employs a sampling approach to systematically generate minimized test suites (i.e., a sampled set of test cases) based on t-way interaction strength. Test suites cover all requisite combinations of parameters and their respective values for tested system configurations, allowing the same tests to perform with fewer test cases than exhaustive testing.

Pairwise testing (also called 2-way testing) is a type of t-way testing technique that is based on the finding that interactions between any two parameter values uncover most faults in the System Under Test (SUT). According to the research conducted by Lockheed Martin and the National Institute of Standard and Technology (NIST) [2], pairwise strategies can detect up to 80 percent of faults while testing various domains of SUT. Each test case in the pairwise test suite covers necessary 2-way parameter combinations at least one time. Thus, it can be regarded as an effective replacement for exhaustive testing.

The primary goal of t-way testing is to produce optimal test cases that cover combinations once at the most. Over the past 20 years, numerous pairwise strategies have been developed, primarily under three approaches: algebraic, pure-computational and metaheuristic. Algebraic strategies are mostly restricted to testing on small system configurations [3] whereas, computational strategies, as deterministic strategies, struggle to generate optimal/near-optimal test suites. Optimization algorithms have started to implement for t-way test data generation as a complement to those two approaches.

Metaheuristic algorithms have been widely applied to solve complex optimization problems in various engineering fields because they offer good results within an acceptable time. Metaheuristic adapted pairwise strategies generally yield the most minimum test cases compared to other approaches. Instances of such strategies include Particle Swarm Optimization (PSO) [4], Ant Colony Optimization (ACO) [5], Harmony Search (HS) [6], Flower Pollination Algorithm (FPA) [7], Cuckoo Search (CS) [8] and Artificial Bee Colony (ABC) [9], to mention a few. Even so, pairwise test generation is an NP-hard problem and no single algorithm can offer optimal solutions for all types of system configurations according to No Free Lunch Theorem (NFL) [10]. Based on the afore-mentioned reasons, we propose a new pairwise strategy adopting GSA, called Pairwise Gravitational Search Algorithm Strategy (PGSAS). In this paper, we discuss the design and implementation of PGSAS. In addition, PGSAS is benchmarked against existing pairwise strategies to demonstrate its performance.

This paper is structured as follows. Section II illustrates the overview of pairwise t-way testing. Section III highlights the related work on developed pairwise strategies. Section IV describes GSA. Section V depicts our proposed PGSAS. Section VI presents the benchmarking results of PGSAS. Finally, Section VII delivers the conclusion.

## II. OVERVIEW OF PAIRWISE TESTING

In order to illustrate the application of pairwise testing, Fig. 1 shows the security surveillance system design as a hypothetical example. The system consists of five input parameters: AC Power Supply, Backup Battery, Surveillance Camera, PIR Motion Detect Sensor and GSM Module. Each parameter takes two selection values. The system parameters and values are shown in Table I.

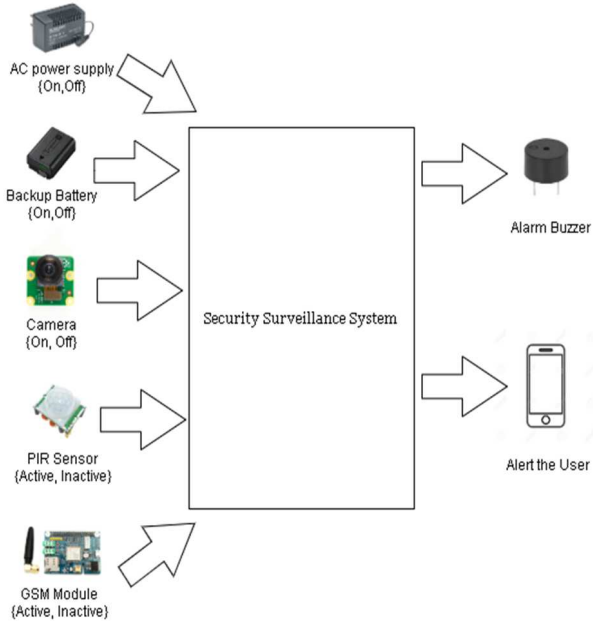

Fig. 1. Security Surveillance System

TABLE I. SYSTEM PARAMETERS AND VALUES

| Parameters | AC Power Supply | Backup Battery | Surveillance Camera | PIR Sensor | GSM Module |
|---|---|---|---|---|---|
| Values | On | On | On | Active | Active |
|  | Off | Off | Off | Inactive | Inactive |

TABLE II. PAIRWISE T-WAY TEST SUITE

| Generated Test Case | AC Power Supply | Backup Battery | Surveillance Camera | PIR Sensor | GSM Module |
|---|---|---|---|---|---|
| 1 | On | On | On | Active | Active |
| 2 | On | Off | Off | Inactive | Inactive |
| 3 | Off | On | On | Active | Inactive |
| 4 | Off | Off | Off | Inactive | Active |
| 5 | On | On | Off | Active | Active |
| 6 | Off | Off | On | Active | Inactive |
| 7 | On | On | On | Inactive | Active |

Testing exhaustively of all possible interactions for the above-mentioned example requires 32 possible test cases (i.e., 2 x 2 x 2 x 2 x 2). Despite the unnoticeable amount of test cases, however, many real-world software systems include enormous parameters and values, resulting in a large test suite. For example, a system with 30 input parameters and only two base values necessitates $2^{20}$ (1048576) exhaustive test lists, directing combinatorial explosion problems [11]. By applying 2-way testing (i.e., pairwise testing), the test suite is reduced from 32 to 7 test cases, as shown in Table II. The generated pairwise test suite covers all required two-parameter combinations and saves 78 percent of the testing time and cost of this SUT example.

## III. RELATED WORK

Among the different approaches used by the developed pairwise t-way strategies, algebraic based strategies are the earliest developed ones. Test Configuration (TConfig) [12] is a notable strategy that uses mathematical algebraic functions to generate pairwise test cases. TConfig supports faster computation time but is designed to test only small SUTs. Pure-computational approach ensures that each generated test case covers the utmost number of uncovered interaction tuples (i.e., parameters and values combinations). In Parameter Order (IPO) implements One-Parameter-at-a-Time (OPAT) method. IPO initially creates a pairwise test set for the first pair of parameters and then increases it by generating three-parameter pairs in horizontal extension until all parameters are covered. This is accompanied by adding test cases in vertical extension to cater for uncovered tuples. IPO is further integrated as the In Parameter Order General (IPOG)[13] strategy and its modified variant strategies. Jenny [14] is an MSDOS-based command-line test suite generator written in C language. Unlike IPO and IPOG, Jenny applies One-Test-at-a-Time (OTAT) method to produce a test case per iteration cycle. A software engineer from Microsoft developed an open-source testing tool called Pairwise Independent Combinatorial Testing (PICT) [15]. PICT also uses OTAT method to generate test cases greedily. Almost all pure-computational strategies are deterministic, i.e., strategies produce the most optimal test size in a single execution. Their search techniques, however, are not as effective as optimization algorithms.

Metaheuristic approach exploits stochastic behavior to search for the best test cases through several iteration cycles. Despite the longer computation time, the performance of the metaheuristic pairwise strategies can be improved by increasing the number of iterations. Pairwise Particle Swarm Based Test Generator (PPSTG) [4] applies Particle Swarm Optimization (PSO) algorithm, which replicates the process of the swarm of birds looking for food. PPSTG initializes a random population of particle swarms (i.e., test cases), as well as the positions and velocities of each particle in the solution search space. At every iteration, the velocity of each test case is updated according to the current best test case attained. Then, the particle shifts to the next better position to search for the next best test case. The process continues until the termination criteria are met, and all test cases are added to the final test suite (FTS).

Pairwise Harmony Search Strategy (PHSS) [6] uses Harmony Search (HS) algorithm to find the best pairwise test cases through a series of improvisation cycles. Harmony Memory (HM), Harmony Memory Consideration Rate (HMCR), Pitch Adjustment Rate (PAR) and Maximum Iteration are the core search control parameters in PHSS.

The cuckoos' unique lifestyle and aggressive reproduction behavior are impersonated in Pairwise Cuckoo Search Strategy (PairCS) [8]. In PairCS, two operations are conducted in each test case generation cycle. A new test case is generated using a Levy flight and compared with the existing test cases. The new test case replaces the current one if it provides a better objective function value. After that, CS adopts a probabilistic elitism operator to maintain elite test cases for the next generation.

Migrating Birds Optimization (MBO) [16] algorithm emulates the energy-saving manner of birds flying in a V shape over a long distance during the winter migration. Two MBO based t-way strategies have been introduced for pairwise test generation. Pairwise MBO strategy (PMBOS) implements original MBO, and Improved Pairwise MBO Strategy (iPMBOS) enhances PMBOS performance by hybridizing with multiple neighborhood structure and elitism operator.

Pairwise Flower Strategy (PairFS) [7] adopts Flower Pollination Algorithm (FPA) as its core implementation. FPA is designed to model the pollination behavior of flowering plants. Three phases are carried out for pairwise test suite production. The first phase sets parameters such as pollen size (population size), switch probability (pa) and stopping criteria. Phase 2 proceeds with the generation of all possible t-way interaction pairs. The final phase produces one complete test case per iteration. Each test case represents a flower or pollen. Optimal test cases obtained at each cycle of iteration are added to FTS, and the covered elements of those test cases are eliminated from the tuple list. The process loop is executed until all elements in the list are covered.

Artificial Bee Colony (ABC) algorithm mimics the way bees search for food sources (nectars) for their beehive. ABC is evolved to solve non-linear optimization problems. Pairwise ABC (PABC) [9] strategy adapts ABC for test suite optimization. There are three types of bees included in the process: Employer bees, Onlooker bees, and Scout bees. The employer bees look for food sources (test cases) and share the information with the onlooker bees. The onlooker bees evaluate the quality of the food source (good test cases) based on the dance performed by the employer bees. Regarding the scout bees, they randomly search for new food sources globally, replacing the abandoned food sources (worse test cases). PABC is a local search-based algorithm since its focus is more concerned with the exploitation search process between employer bees and onlooker bees.

The novel Kidney Algorithm (KA) influences the evolution of Pairwise Kidney Strategy (PKS) [17]. KA imitates kidney processes in the human body: filtration, reabsorption, secretion, and excretion. Filtration acts as a local search to filtrate the generated solutions (test cases) into filtered blood (FB) (i.e., good solutions) group and waste group (W) (i.e., worst solutions). The reabsorption process acts like a global search where the test cases in W are re-evaluated to be redirected back to FB. In the process of secretion, added test cases in FB are checked. The test cases with inadequate quality coverage are brought back to W. Finally, the test cases in W are eliminated and replaced with newly generated test cases.

Dragonfly Optimization Algorithm (DFA) [18] based t-way strategy is motivated by the two ways used by dragonflies swarms called static and dynamic. Dragonflies statically separate themselves into small groups to fly in a small area to look for food sources such as butterflies and various insects, implementing the global search phase. On the other hand, they dynamically fly in bigger groups in a single direction, undertaking the local search phase. Each dragonfly (i.e., a test case) is assessed based upon its quality of food source (i.e., interaction coverages), and the same goes for its nearby dragonflies. Selected best dragonfly (i.e., best test case covering most uncovered interaction elements) is added to FTS and covered elements are removed from the list.

IV. GRAVITATIONAL SEARCH ALGORITHM

Gravitational Search Algorithm (GSA) [19] was developed in 2009. Since then, GSA has piqued the interest of research in various engineering fields due to its efficiency in solving non-linear optimization problems. GSA is a metaheuristic optimization algorithm based on two well-known laws of Issac Newton: the law of universal gravitation and the law of motion interaction. In GSA, it is assumed that a hypothetical universe consists of a population of objects (search space) heading towards each other because of gravitational attraction force. Each object has its own position vector in certain dimensions, designating as potential solutions to an optimization problem. Objects' performances are assessed by their respective masses that are assigned with fitness function values. Objects with heavier masses and higher fitness values (good solutions) move slower than the lighter ones (worse solutions). The position and velocity of each individual object are updated at every iteration process, and the best fitness value, along with its correlated object, is reserved. After several iterations, the object with the heaviest mass (optimal solution) is aimed to attract the whole collection of other objects in the search space. The above-stated procedures of GSA are iterated until the stopping conditions are encountered. The detailed processes of GSA are described in Steps A through G.

*A. Search Space Initialization*

GSA initializes population of objects in the search space. Number of solutions S with n-dimensional solution spaces is declared as follows:

$$X_i = (x_i^1, ..., x_i^d, ..., x_i^n) \text{ for } i=1,2,...,S \qquad (1)$$

Where $X_i$ represents the position vector of ith object and $x_i^d$ is the position of ith object at dth dimension.

*B. Fitness Evaluation of Objects*

The best and worst fitness values of objects are calculated for either a maximization or minimization problem. For maximum oriented optimization, Equation (2) and (3) are used, and Equation (4) and (5) are defined respectively for minimum oriented optimization.

$$best = max_{i \in 1,...,S}[fit(X_i)] \qquad (2)$$

$$worst = min_{i \in 1,...,S}[fit(X_i)] \qquad (3)$$

$$best = min_{i \in 1,...,S}[fit(X_i)] \qquad (4)$$

$$worst = max_{i \in 1,...,S}[fit(X_i)] \qquad (5)$$

Where *best and worst* represent the best and the worst fitness values in the population and *fit(X_i)* represents the fitness value of ith object.

## C. Mass Calculation

The performance measure of each object is computed as in Equation (6). To have smoother simulation process, the mass values are normalized between 0 and 1 and is expressed as in Equation (7).

$$m(X_i) = \frac{fit(X_i) - worst}{best - worst} \quad (6)$$

$$M(X_i) = \frac{m(X_i)}{\sum_{i=1}^{S} m(X_i)} \quad (7)$$

Where $m(X_i)$ is the mass of ith object and $M(X_i)$ is the relative normalized mass of ith object.

## D. Gravitational Constant Calculation

Gravitational constant G is one of the core parameters of GSA that controls exploration search process accuracy. Equation (8) describes the formula of G at time t, G(t).

$$G(t) = G_0 \times e^{-\alpha \frac{t}{T_{max}}} \quad (8)$$

Where $G_0$ represents the initial $G(t)$ value, $\alpha$ is the arbitrary parameter that controls GSA convergence rate, $t$ is the current iteration step and $T_{max}$ is the maximum iteration.

## E. Gravity Force Calculation

$$F_{ij}^d = G \frac{M(X_i) \times M(X_j)}{R_{ij} + \varepsilon} [x_j^d - x_i^d] \quad (9)$$

Where $F_{ij}^d$ represents the force exerting from jth object on ith object at dth dimension, $R_{ij}$ is the Euclidean distance between ith and jth object, $\varepsilon$ is the small constant to disallow division by zero error and $x_j^d$ and $x_i^d$ are jth and ith objects' positions at dth dimension. $R_{ij}$ is calculated as follows:

$$R_{ij} = \|M(X_j) - M(X_i)\|_2 \quad (10)$$

The total force acting on ith object at dth dimension from other objects is computed using Equation (11).

$$F_d^i = \sum_{j \in K_{best}, j \neq i}^{S} rand_j F_{ij} \quad (11)$$

Where $K_{best}$ represents the set of first K objects with the best fitness value and heaviest mass that will only apply force to others, and *rand* is the random number in the interval [0,1].

## F. Acceleration and Velocity Calculation

Obeying Newton's second law, GSA calculates the movement of objects because of exerted gravity force. The acceleration of any object is directly proportional to the force acting upon that object and inversely proportional to its normalized mass value. Equation (12) provides the formula for acceleration.

$$a_i^d = \frac{F_i^d}{M(X_i)} \quad (12)$$

Where $a_i^d$ represents the acceleration of ith object at dth dimension and $M(X_i)$ refers to the relative normalized mass of ith object. The new velocity for the next iteration is computed as follows:

$$v_i^d(t+1) = rand \times v_i^d + a_i^d \quad (13)$$

Where $v_i^d(t+1)$ represents the new velocity of ith object at dth direction and $v_i^d$ is the previous velocity of ith object at dth dimension.

## G. Updating Objects' Position

Using Equation (14), the position of objects at next iteration is computed.

$$x_i^d(t+1) = x_i^d + v_i^d(t+1) \quad (14)$$

Where $x_i^d(t+1)$ is the new position and $x_i^d$ is the position of ith object at previous iteration. Steps B to G are iterated until the stopping criteria are reached.

## V. PAIRWISE GRAVITATIONAL SEARCH ALGORITHM STRATEGY

A proposed new strategy known as Pairwise Gravitational Search Algorithm (PGSAS) is presented in this section for generating optimal/near-optimal pairwise t-way test suite. Three main stages are involved in this strategy: Stage 1: Input SUT Analysis and Parameter Initialization, Stage 2: Generation of Interaction Tuples, and Stage3: Generation of Pairwise Test Suite.

In Stage 1, PGSAS acquires the input parameters p and values v of SUT. It also initializes the necessary control parameters of GSA particularly taken into consideration in this proposed strategy which are Object Population (N), Initial Gravitational Constant ($G_0$), attenuation factor ($\alpha$), epsilon ($\varepsilon$) and total iteration (T).

In the second stage, PGSAS generates all possible paired parameter combinations. Referring to the example system in Fig. 1, for five input parameters (i.e., A, B, C, D and E are used for simplification purposes), there are ten possible parameter pairs combinations: AB, AC, AD, AE, BC, BD, BE, CD, CE and DE. Based on the combination result, the interaction tuples for all parameters are generated and added to the Interaction Tuple List (ITL) accordingly. Since two base values are contributed in each parameter of SUT, there will be 2x2 interaction tuples for every parameter combination. For instance, the combination BD parameters will have 2x2 tuples of b0:d0, b0:d1, b1:d0, b1:d1 considering B parameter has b0 and b1 values and D parameter has d0 and d1 values. The same process is repeated for the remaining parameter combinations.

The third stage is carried out using GSA to search for test cases that cover the maximum number of tuples from ITL. PGSAS randomly sets a population of objects and generates a single test case at every iteration cycle, following OTAT method. The weight coverage function evaluates the quality of test cases. The weight or the fitness is the number of interaction tuples that can be covered by the candidate test case, which is computed as follows:

$$fit(X_i) = \sum_{i=0}^{tc} w_i \qquad (15)$$

Where $fit(X_i)$ represents the fitness function of the candidate test case $X_i$, $i$ is the interactions, $tc$ is the total number of interaction tuples in the ITL and $w_i$ is the number of uncovered tuples (weight coverage) from ith interaction. Among the generated test cases, the test case with the maximum weight coverage (i.e., covering most uncovered tuples from ITL) is chosen and added to FTS. After that, the covered tuples by the test case are eliminated from ITL. For the next iteration, the new position of the object is updated (i.e., new test case) based on its current position and new velocity using Equation (14). The process goes on until the designated number of iterations is ended or until the best test case can no longer be found from ILT (i.e., ILT becomes empty). Fig. 2 depicts the pseudocode of Pairwise Gravitational Search Algorithm Strategy.

| 01: | **Input**: Parameters (p) and Values (v) of System Under Test (SUT) |
| 02: | : Parameters of GSA: $G_0$, α, ε and T |
| 03: | **Output**: Final Pairwise Test Suite (FTS) |
| 04: | **Begin** |
| 05: | Generate all possible p-combinations with 2-way strength (t = 2) |
| 06: | Generate interaction tuples and add to Interaction Tuple List (ITL) |
| 07: | **while** ITL is not empty **do** |
| 08: |    Set objects' population (N) randomly |
| 09: |    **for** current iteration (t) <= Total Iteration (T) |
| 10: |       Evaluate fitness value for each object $fit(X_i)$ to check weight coverage from ITL using Eq. (15) |
| 11: |       Evaluate best and worst fitness values in N using using Eq. (4) and (5) |
| 12: |       **if** best != worst |
| 13: |          Calculate $m(X_i)$ and $M(X_i)$ by Eq. (6) and (7) |
| 14: |          Calculate $F_d^i$ and $a_i^d$ using Equations from steps E and F in Section IV |
| 15: |          Update $V_i$ using Eq. (13) |
| 16: |          Update $X_i$ using Eq. (14) |
| 17: |       **end if** |
| 18: |    **end for** |
| 19: |    Evaluate the best test case $X_{best}$ |
| 20: |    Add $X_{best}$ into FTS |
| 21: |    Eliminate covered interaction tuples from ITL |
| 22: | **end while** |
| 23: | **End-Process** |

Fig. 2. Pairwise Gravitational Search Algorithm Pseudocode

## VI. RESULTS AND DISCUSSION

In this section, we compare our proposed PGSAS to other existing pairwise t-way strategies in the literature to assess its performance in terms of the size of the produced pairwise test suite. Prior to benchmarking experiments, parameter tuning process is first conducted for improving the performance of PGSAS. Four parameters of GSA (population size N, initial gravitational constant $G_0$, attenuation factor α and total iteration T) are tuned with different range of values. The optimal results are obtained when N = 200, $G_0$ = 10, α = 20 and T = 500. Thus, we adopt these parameter values as base values for our experiments. Since PGSAS is a stochastic strategy, we execute it 30 times for every benchmarking configurations and record only the best test suite size. We use Java JDK 13.0.1 for developing and experimenting PGSAS. The experiments are divided into two groups. In the first group, we compare PGSAS with published results of existing pairwise strategies using eleven various system configurations. For example, SC1: $2^7$ refers to a configuration with 7 parameters 2-valued each. As for the second group, the comparison is done between PGSAS and other strategies by using a system configuration with v = 2 and p parameters are varied from 3 to 15, with an additional 50 for testing higher configurations.

Table III and IV show the results of all benchmarking experiments. The best test suite sizes obtained by each strategy are marked with bold numbers in the table cells. Cells marked as "NA" denote that the results are not available in the published papers.

Regarding the results reported in Table III, PGSAS produces optimal results in six configurations which are SC1, SC2, SC4, SC5, SC6 and SC7 but does not outperform other strategies. PairCS generates the best result in SC5, whereas PairFS produces its optimal result in SC10 and SC11. PKS, on the other hand, delivers the smallest test suite size in SC8, while PPSTG and DFA provide their best results in SC9. Although PGSAS performs poorly in the remaining configurations, it still generates acceptable test sizes compared to existing strategies.

Referring to Table IV, PGSAS and PKS achieve optimum results among all system configurations of p variants ranging from 3 to 15, outshining all other strategies. However, PGSAS generates the most minimum pairwise test suite in the configuration with p = 8. The rest of the strategies often obtain competitive outcomes, with similar or identical results to those of PGSAS and PKS. PGSAS also expresses the potential to address higher system configurations by testing a p value of 50, and the produced result is comparable to that of PairCS. Overall, PGSAS gives the most optimal results for smaller system configurations, while for larger configurations, it generates either better or comparable results to existing strategies.

## VII. CONCLUSION

This paper presents our proposed method called Pairwise Gravitational Search Algorithm Strategy (PGSAS) for pairwise test suite generation. Our benchmarking results demonstrate that PGSAS tends to generate one optimal pairwise test case per iteration cycle and obtains competitive results in most cases. From the results, it can also be seen that none of the strategies manages to produce the best test size for all types of configurations (NP-hard problem). PGSAS bridges the gap by delivering optimal results in configurations where other strategies cannot provide. However, GSA, like most other metaheuristic algorithms, struggles from premature convergence. Thus, we plan to hybridize other optimization algorithms with PGSAS to improve the current performance. We are also working to support uniform t-way testing with higher interaction strength (t > 2), as well as variable strength interaction testing.

TABLE III. COMPARISON OF PGSAS WITH EXISTING STRATEGIES USING DIFFERENT SYSTEM CONFIGURATIONS

| System Configurations (SC) | TConfig | Jenny | PICT | IPOG | PPSTG | PHSS | PairCS | PairFS | PABC | PKS | DFA | PGSAS |
|---|---|---|---|---|---|---|---|---|---|---|---|---|
| SC1: $2^7$ | 7 | 8 | 7 | 7 | **6** | NA | **6** | NA | NA | NA | NA | **6** |
| SC2: $3^7$ | 15 | 16 | 16 | **15** | **15** | NA | **15** | NA | **15** | NA | NA | **15** |
| SC3: $4^7$ | 28 | 28 | 27 | 29 | 26 | NA | **25** | NA | NA | NA | NA | 26 |
| SC4: $3^3$ | 10 | 10 | 10 | 11 | **9** | 9 | 9 | 9 | 9 | 9 | 9 | 9 |
| SC5: $3^4$ | 10 | 13 | 13 | 12 | **9** | 9 | 9 | 9 | 9 | 9 | 9 | 9 |
| SC6: $3^5$ | 14 | 14 | 13 | 15 | 12 | NA | **11** | NA | NA | NA | NA | **11** |
| SC7: $2^{10}$ | 9 | 10 | NA | NA | **8** | NA | **8** | NA | NA | NA | NA | **8** |
| SC8: $3^{10}$ | 17 | 19 | 18 | 20 | NA | 17 | NA | NA | 17 | **16** | 17 | 17 |
| SC9: $3^{13}$ | 20 | 22 | 20 | 20 | **17** | 18 | 18 | 18 | 18 | 20 | **17** | 20 |
| SC10: $4^{10}$ | 31 | 30 | 31 | 31 | NA | 29 | NA | **28** | **28** | 30 | 30 | 31 |
| SC11: $5^{10}$ | 48 | 45 | 47 | 50 | NA | 45 | NA | **42** | 43 | 46 | 45 | 48 |

TABLE IV. COMPARISON OF PGSAS WITH EXISTING STRATEGIES USING V=2 AND P VARIED FROM 3 TO 50

| P | TConfig | Jenny | PICT | IPOG | PPSTG | PHSS | iPMBOS | PairCS | PairFS | PABC | PKS | PGSAS |
|---|---|---|---|---|---|---|---|---|---|---|---|---|
| 3 | **4** | 5 | **4** | **4** | **4** | **4** | **4** | **4** | **4** | **4** | **4** | **4** |
| 4 | 6 | 6 | **5** | 6 | 6 | 6 | 6 | **5** | 6 | **5** | **5** | **5** |
| 5 | **6** | 7 | 7 | **6** | **6** | **6** | **6** | **6** | **6** | **6** | **6** | **6** |
| 6 | 7 | 8 | **6** | 8 | 7 | 7 | 7 | **6** | 7 | 7 | **6** | **6** |
| 7 | 9 | 8 | 7 | 8 | 7 | 7 | 7 | 7 | 7 | 7 | **6** | **6** |
| 8 | 9 | 8 | 8 | 8 | 8 | 8 | 7 | 8 | 8 | 8 | 7 | **6** |
| 9 | 9 | **8** | 9 | **8** | **8** | **8** | **8** | **8** | **8** | **8** | **8** | **8** |
| 10 | 9 | 10 | 9 | 10 | **8** | **8** | **8** | **8** | **8** | **8** | **8** | **8** |
| 11 | 9 | 9 | 9 | 10 | 9 | **8** | **8** | **8** | **8** | **8** | **8** | **8** |
| 12 | 9 | 10 | 9 | 10 | 9 | 9 | **8** | 9 | 9 | 9 | **8** | **8** |
| 13 | NA | 10 | 9 | 10 | 9 | 9 | 9 | NA | NA | 9 | **8** | **8** |
| 14 | NA | 10 | 10 | 10 | 9 | 10 | **9** | NA | NA | **9** | **9** | **9** |
| 15 | NA | 10 | 10 | 10 | 10 | 10 | **9** | NA | NA | **9** | **9** | **9** |
| 50 | NA | NA | NA | NA | NA | NA | NA | **12** | NA | NA | NA | 13 |

ACKNOWLEDGMENT

The authors would like to acknowledge the support from Fundamental Research Grant Scheme (FRGS) under a grant number of FRGS/1/2018/ICT01/UNIMAP/02/1 from the Ministry of Education Malaysia.

REFERENCES

[1] B. S. Ahmed, A. Gargantini, K. Z. Zamli, C. Yilmaz, M. Bures, and M. Szeles, "Code-Aware Combinatorial Interaction Testing," 2019.

[2] L. Hu, W. E. Wong, D. R. Kuhn, and R. N. Kacker, "How does combinatorial testing perform in the real world: an empirical study," *Empir. Softw. Eng.*, vol. 25, no. 4, pp. 2661–2693, 2020.

[3] A. K. Alazzawi, H. Md Rais, and S. Basri, "Artificial Bee Colony Algorithm for t-Way Test Suite Generation," in *2018 4th Int. Conf. on Comput. Inf. Sci. Revolutionising Digit. Landsc. Sustain. Smart Soc. ICCOINS 2018 - Proceedings*, 2018.

[4] B. S. Ahmed and K. Z. Zamli, "The development of a particle swarm based optimization strategy for pairwise testing," *J. Artif. Intell.*, vol. 4, no. 2, pp. 156–165, 2011.

[5] N. Ramli, R. R. Othman, and R. Hendradi, "A uniform strength t-way test suite generator based on ant colony optimization algorithm to produce minimum test suite size," *AIP Conf. Proc.*, vol. 2339, no. 1, 2021.

[6] A. R. A. Alsewari and K. Z. Zamli, "A harmony search based pairwise sampling strategy for combinatorial testing," *Int. J. Phys. Sci.*, vol. 7, no. 7, pp. 1062–1072, 2012.

[7] A. B. Nasser, A. R. A. Alsewari, N. M. Tairan, and K. Z. Zamli, "Pairwise test data generation based on flower pollination algorithm," *Malaysian J. Comput. Sci.*, vol. 30, no. 3, pp. 242–257, 2017.

[8] A. B. Nasser, Y. A. Sariera, A. R. Alsewari, and K. Z. Zamli, "A Cuckoo Search Based Pairwise Strategy for Combinatorial Testing Problem," vol. 82, no. 1, 2015.

[9] A. K. Alazzawi, A. A. Ba Homaid, A. A. Alomoush, and A. R. A. Alsewari, "Artificial Bee Colony algorithm for pairwise test generation," *J. Telecommun. Electron. Comput. Eng.*, vol. 9, no. 1–2, pp. 103–108, 2017.

[10] S. P. Adam, S. A. N. Alexandropoulos, P. M. Pardalos, and M. N. Vrahatis, "No free lunch theorem: A review," *Springer Optim. Its Appl.*, vol. 145, pp. 57–82, 2019.

[11] K. Z. Zamli, F. Din, B. S. Ahmed, and M. Bures, "A hybrid Q-learning sine-cosine-based strategy for addressing the combinatorial test suite minimization problem," *PLoS One*, vol. 13, no. 5, pp. 1–29, 2018.

[12] A. W. Williams, "Determination of Test Configurations for Pair-wise Interaction Coverage," *Test. Commun. Syst.*, 2000.

[13] R. N. Kacker, D. R. Kuhn, Y. Lei, and D. E. Simos, "Factorials Experiments, Covering Arrays, and Combinatorial Testing," *Math. Comput. Sci.*, no. April, 2021.

[14] B. Jenkins, "Jenny." http://www.burtleburtle.net/bob/math/jenny.html (accessed Feb. 13, 2020).

[15] J. Czerwonka, "Pairwise Testing in Real World: Practical Extensions to Test Case Generator," *Proc. 24th Pacific Northwest Softw. Qual. Conf.*, pp. 419–430, 2006.

[16] H. L. Zakaria and K. Z. Zamli, "Migrating Birds Optimization based strategies for Pairwise testing," *2015 9th Malaysian Softw. Eng. Conf. MySEC 2015*, pp. 19–24, 2016.

[17] A. A. B. Homaid, A. A. Alsewari, A. K. Alazzawi, and K. Z. Zamli, "A Kidney Algorithm for Pairwise Test Suite Generation," *Adv. Sci. Lett.*, vol. 24, no. 10, pp. 7284–7289, 2018.

[18] B. S. Ahmed, "Generating pairwise combinatorial interaction test suites using single objective dragonfly optimisation algorithm," *arXiv*, 2019.

[19] E. Rashedi, H. Nezamabadi-pour, and S. Saryazdi, "GSA: A Gravitational Search Algorithm," *Inf. Sci. (Ny).*, vol. 179, no. 13, pp. 2232–2248, 2009.